# Simulation of cascades caused by UHE and EHE neutrinos in dense media

Igor Zheleznykh[1], Leonid Dedenko[2], Grigorii Dedenko[3], Anna Mironovich[1]

1. Institute for Nuclear Research of Russian Academy of Sciences, Moscow, 117312, 60th Anniversary of October Prospect, 7a, Russian Federation
2. Moscow State University, Faculty of Physics, Moscow, 119992, Leninskie Gory, Russian Federation
3. Federal State Autonomic Educational Institution of Higher Professional Education, National Research Nuclear University «MEPhI», Moscow, 115409, Kashirskoye shosse 31, Russian Federation

**Abstract:** A method of simulation of particle cascades induced by ultra-high ($>10^{15}$ eV) and extremely high ($>10^{18}$ eV) energy neutrinos in water or other dense medium has been elaborated. The lateral spread of high-energy particles in cascades due to Coulomb scattering is negligible. So it is possible to use an approximation of the one-dimensional development of the UHE (or EHE) cascade if energies of particles in it higher than 10-100 GeV. An original program of the one-dimensional development of UHE and EHE cascades in dense media taking into account fluctuations and the Landau-Pomeranchuk-Migdal effect has been elaborated. The GEANT4 package has been used when particle energies are below 1000 GeV, for example, or less to calculate correctly the longitudinal distribution of an energy deposition by charge particles in a cascade. In advance the library of longitudinal characteristics of cascades was calculated. When a particle with the energy E which is below the threshold energy (E<1000 GeV or E<$10^6$ GeV) appears in the basic Monte Carlo program, the longitudinal distribution of an energy deposition induced by it was taken from the library and was added into the corresponding places of the basic distribution. The results of simulation of cascades in a dense medium can be used for calculations of acoustical or radio signals.

# 1. Introduction. Alternative KM3 neutrino telescopes and simulations of UHE and EHE particle cascades

The development of methods of the high-energy (HE) cosmic neutrino detection, which was, initiated my M. Markov [1], F. Reines [2], K. Greisen [3] in the end of $50^{th}$ – the beginning of $60^{th}$ is one of the main branches of the Cosmic Rays Physics. The ideas to detect HE atmospheric neutrinos and to study neutrino interaction in the deep underground and deep underwater neutrino experiments were supplemented with the suggestion to carry out searches for HE neutrinos (as well as HE gamma quanta) from Outer Space, from such astrophysical objects as supernova remnants (Crab Nebula, for example), the center of our Galaxy etc. [4, 5]. In fact the matter was about a new branch of the Astronomy also: to investigate the Universe using HE neutrino detectors – neutrino telescopes. However the detection of weakly interacting cosmic neutrinos even of ultrahigh energy (UHE, >$10^{15}$ eV) and extremely high energy (EHE, >$10^{18}$ eV) is very difficult task. It took 50 years to develop new methods of cosmic neutrino detection and to construct and to suggest a number of underground, underwater, under ice (RAMAND, IceCube, RICE, ARIANNA, ARA) and even under moon (radio astronomical RAMHAND-type: GLUE, Kalyazin, NuMoon, LUNASKA) large-scale neutrino telescopes. The surface (EAS) detectors of Auger-type, ANITA (using stratosphere balloon) and future orbital detectors of JEM-EUSO-type have to be included in this list also. The careful description and analysis of performed investigations with cosmic neutrinos and perspectives of the UHE and EHE neutrino investigations had been made in the review of Christian Spiering "Towards high energy neutrino astronomy" [6], see also References there.

The great contribution of the DUMAND project (F. Reines, H. Bradner, A. Roberts, J. Learned, V. Stenger, M. Shapiro, G. Wilkins et al) since 1975 and DUMAND Hawaii center (since 1980) to the development of the method of the deep underwater optical neutrino detection and the scientific program for neutrino telescopes of cubic kilometer size (KM3) was emphasized in [6]. Besides in the framework of the DUMAND the first alternative (hydro-acoustical) method of UHE neutrino detection began to develop and the first calculations of acoustical pulses from particle cascades produced by UHE neutrinos in the Ocean had been made.

The DUMAND project stimulated the wide interest to the deep underwater investigations for neutrino astronomy in the Soviet Union. In the framework of the Soviet DUMAND program (1981-1991) under leadership of M. Markov the Baykal Lake and the Mediterranean Sea were investigated as neutrino targets, the R&D for the Baykal neutrino telescope was carried out, a few prototype modules for the deep underwater neutrino telescope NESTOR and the hydro-acoustic neutrino telescope SADCO were designed and tested in the Mediterranean Sea. But two perspective neutrino targets - Antarctica and the Moon were also considered and new alternative detection methods for UHE and EHE neutrino astronomy were suggested: Radio Antarctic Muon And Neutrino Detection - RAMAND (1983) and Radio Astronomical Method of Hadron And Neutrino Detection - RAMHAND (1988-1989). Four expeditions to Antarctic station Vostok in 1985-1990 had shown good perspectives for development of radio wave KM3 neutrino telescopes in Antarctica.

At the same time the new calculation methods [7-20] have been elaborated to simulate development of the UHE and EHE cascades in dense media with the LPM effect taken into account [21, 22]. These cascades are supposed to be detected with the help of the KM3 neutrino telescopes. First, the Greisen (NKG) formulae [15] are suggested to be used to simulate a development of a cascade at low energies. The original scheme to take into account the LPM effect at high energies has been proposed [7,8]. Later the GEANT4 package has been proposed to be used [20].

In this paper the longitudinal distributions of energy depositions in the electron-photon cascades in the water have been estimated taking into account the LPM effect. First, the library of longitudinal distributions of an energy deposition dE/dZ in the low energy cascades induced by gammas in water have been simulated with the help of the Geant4 package [23] and the original code which takes into account the LPM effect [21]. Then calculations of longitudinal distributions of energy depositions in the high energy cascades in water were carried out with the help of the simulated library and the original code. This method was called as the hybrid LPM-G4 algorithm.

## 2. Calculation technique

*2.1 Calculations a library of the low energy cascades.*

The longitudinal distributions of the energy deposition dE/dZ in water have been calculated with the help of the Geant4 package [23] for gammas with energies of 0.1, 1, 10, $10^2$ and $10^3$ GeV with statistics $10^6$, $10^5$, $10^4$, $10^3$ and $10^2$ respectively. The energy depositions dE/dZ have been simulated in a cylinder of water with the length of 10 m with the step of 1 cm. Results of calculations are shown in Figure 1.

The main parameters of these cascades are shown in Table 1. This Table 1 shows the positions $Z_{max}$ of the maximum of energy deposition dE/dZ distributions and the effective lengths $Z_0$ of the electron-photon cascades in water calculated at 5% of the maximum values of dE/dZ for different energies of gammas.

**Table 1.** Positions $Z_{max}$ of the maximum of energy deposition dE/dZ distributions and the effective lengths $Z_0$ of the electron-photon cascades in the water.

| $E_0$ (GeV) | $Z_{max}$ (cm) | $Z_0$ (m) |
|---|---|---|
| $10^{-1}$ | ~ 29.5 | ~ 2.4 |
| $10^0$ | ~ 92.2 | ~ 3.8 |
| $10^1$ | ~ 178.8 | ~ 5.2 |
| $10^2$ | ~ 263.9 | ~ 6.3 |
| $10^3$ | ~ 334.9 | ~ 6.7 |

These cascades were used to calculate the tail-showers in the frame of the hybrid LPM-G4 method as follows. The development of a cascade of particles with energies above the threshold value $E_{th}$ in water was simulated by the Monte Carlo method with the Migdal cross sections [22] taken into account the LPM effect. If a particle with the energy E below this the threshold value $E_{th}$ appears in this cascade then a longitudinal distribution of an energy deposition of this tail-cascade adjusted to the energy E with the help of the linear approximation was added at appropriate places of longitudinal distribution of an energy distribution of the main cascade. First, the value of the threshold energy was put to $10^3$ GeV to calculate additional library longitudinal distributions of an energy deposition for gammas with energies $10^4$, $10^5$ and $10^6$ GeV with the help of our original code and the GEANT4 package. Then this value was set to $E_{th}=10^6$ GeV and simulations of the longitudinal distributions of an energy deposition in cascades in water induced by gammas with energies $10^7$, $10^8$, $10^9$, $10^{10}$ and $10^{11}$ GeV have been carried out.

*2.2 Calculation of low energy cascades by the hybrid LPM-G4 algorithm for energies $10^4$ - $10^6$ GeV.*

The calculations of the longitudinal distributions of an energy deposition in cascades of particles in water induced by gammas with energies $10^4$, $10^5$ and $10^6$ GeV have been carried out by the hybrid LPM-G4 algorithm. As the "tail" longitudinal distributions for cascades induced by particles with energies E below the threshold $E_{th}=10^3$ GeV a linear approximation of date from the library longitudinal distributions has been used. These longitudinal distributions of the energy deposition dE/dZ in cascades induced in water by gammas with energies $10^4$, $10^5$, $10^6$ GeV are shown in Figure 2. They are used as the additional library distributions.

### 3. Results and discussions

The calculated library of the longitudinal distributions of the energy deposition in cascades induced by particles with energies below the threshold energy $E_{th}=10^6$ GeV allows to simulate the longitudinal distributions of the energy deposition dE/dZ in cascades induced in water by gammas with ultra high energies $10^7$, $10^8$, $10^9$, $10^{10}$ and $10^{11}$ GeV with the help of the hybrid LPM-G4 algorithm. For the longitudinal distributions of the energy deposition in "tail" cascades induced by particles with energies below $E_{th}=10^6$ GeV a linear approximation of the 8 library distributions in the energy range of $10^{-1}$-$10^6$ GeV was used. These "tail" distributions have been added to the main longitudinal distribution at the right places. The longitudinal distributions of the energy deposition dE/dZ in cascades induced in water by gammas with ultra high energies $10^7$, $10^8$, $10^9$, $10^{10}$ and $10^{11}$ GeV are shown in Figure 3. It should be mentioned that the LPM effect is of the primary importance for development of cascades with energies above $10^6$ GeV. As it is seen from Figure 3 such values as a position $Z_{max}$ of the maximum of the longitudinal distribution is loosing its sense. The distributions became looking rather as a "step". Other even more important point is tremendous fluctuations in the development of cascades in water. The longitudinal distributions do not look like a good curve due to these fluctuations. They are rather the random curves.

Figure 4 illustrates clearly the importance of fluctuations. Such definition as an average cascade has no sense if statistics is not too high.

The positions $Z_{max}$ of the maximum of the energy deposition dE/dZ in cascades induced in water by gammas with energies in the range of $10^{-1}$ - $10^{11}$ 10 GeV are presented in Figure 5. The tremendous increase of the positions $Z_{max}$ at the ultra high energies is accounted due to the LPM effect.

To illustrate the role of fluctuations in a development of electron-photon cascades in water due to the LPM effect some additional simulations of the longitudinal distributions of the energy deposition have been carried out to estimate the average distribution for statistics of 10 events and four individual distributions.

Figure 6 shows this average distribution for 10 events (a) and four individual distributions (b, c, d, e, f) for cascades induced in water by gammas with the energy of $10^{11}$ GeV. It is seen that in this case the "average" distribution has not much sense. So, fluctuations in a development of a cascade in water induced by gammas with ultra high energies are very huge.

## 4. Conclusion

In the summary report at the ARENA-2005 conference [24] John Learned noticed the importance to have sufficiently detailed and grounded simulation programs for the UHE neutrino astronomy. In particular the simulation programs of UHE and EHE cascades in solids were mentioned. He emphasized also that the great progress had been made in using GEANT and CORSIKA for this goal and that a hybrid approach suggested in [20] should also help in this area.

Since 2005 a number of codes have been developed to simulate UHE cascades as well as acoustic and radio signals from cascades in solids, see for example [25-29].

The very economical and reasonable scheme has been suggested above to simulate a development of the electron-photon cascades in dense media at ultra-high energies with the LPM effect taken into account. The hybrid LPM-G4 algorithm used as a base of this scheme includes simulations of the tree-dimensional electron-photon cascades in dense media at low energies with the help of the GEANT4 package and calculations of the one-dimensional development of these cascades at ultra-high energies with the LPM effect taken into account with the help of our original code. A statistics of $10^5$-$10^6$ events at low energies provides a good accuracy of simulations. Besides the GEANT4 package guarantees the correct lateral spread of charge particles due to the Coulomb scattering. In our previous papers we have used the modified NKG function to take into account this scattering. At last, our original code takes very reasonably the LPM effect into account. So, the results for the distributions of an energy deposition in the electron-photon cascades presented in this paper are expected to be quite reasonable.

## Acknowledgement

The authors are deeply indebted to David Besson for valuable discussions.

**Figures captions**

**Fig. 1.** The longitudinal distributions of the energy deposition dE/dZ in cascades induced in water by gammas with energies of 0.1, 1, 10, $10^2$ and $10^3$ GeV

**Fig. 2.** The longitudinal distributions of the energy deposition dE/dZ in cascades induced in water by gammas with energies $10^4$, $10^5$, $10^6$ GeV

**Fig. 3.** The longitudinal distributions of the energy deposition dE/dZ in cascades induced in water by gammas with energies $10^7$, $10^8$, $10^9$, $10^{10}$ and $10^{11}$ GeV

**Fig. 4.** The longitudinal distributions of the energy deposition dE/dZ in cascades induce in water by gammas with energies $10^{10}$ and $10^{11}$ GeV.

**Fig. 5.** The position $Z_{max}$ of the maximum of the longitudinal distributions of the energy deposition dE/dZ in cascades induced in water by gammas with energies in the range of $0.1$-$10^{11}$ GeV.

**Fig. 6.** The longitudinal distributions of the energy deposition dE/dZ in cascades induced in water by gammas with energy of $10^{11}$ GeV: (a) – the average distribution for 10 events, (b, c, d, e, f) – individual distributions.

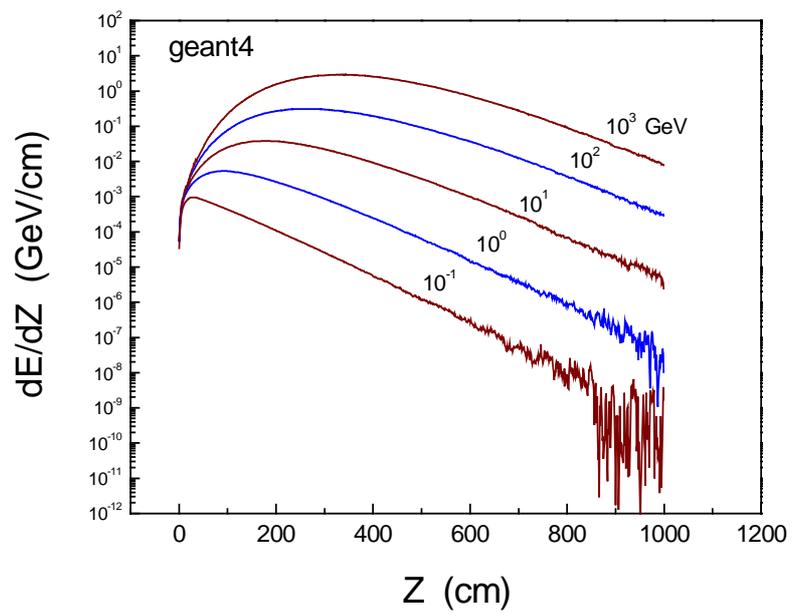

**Fig. 1.**

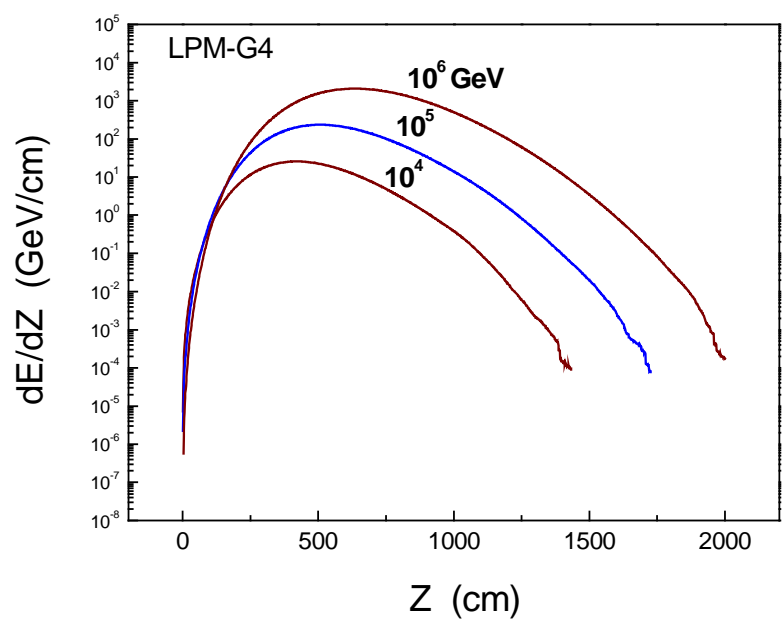

**Fig. 2.**

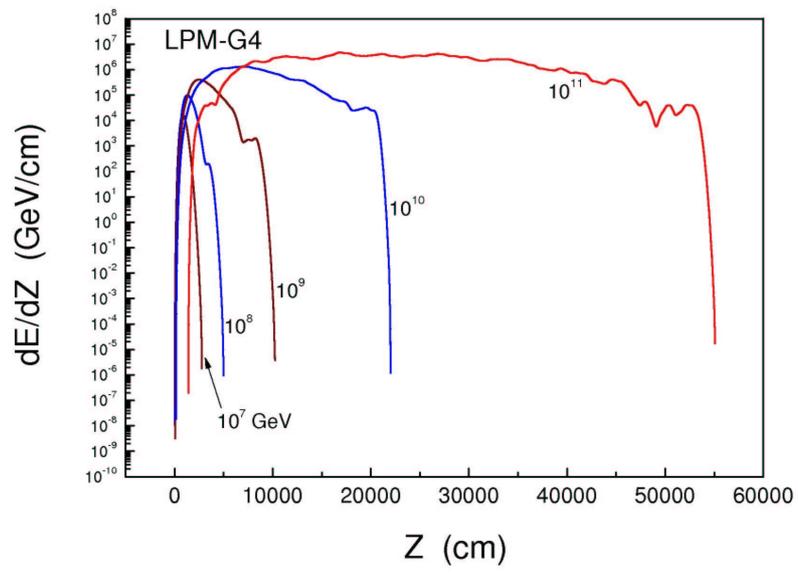

**Fig. 3.**

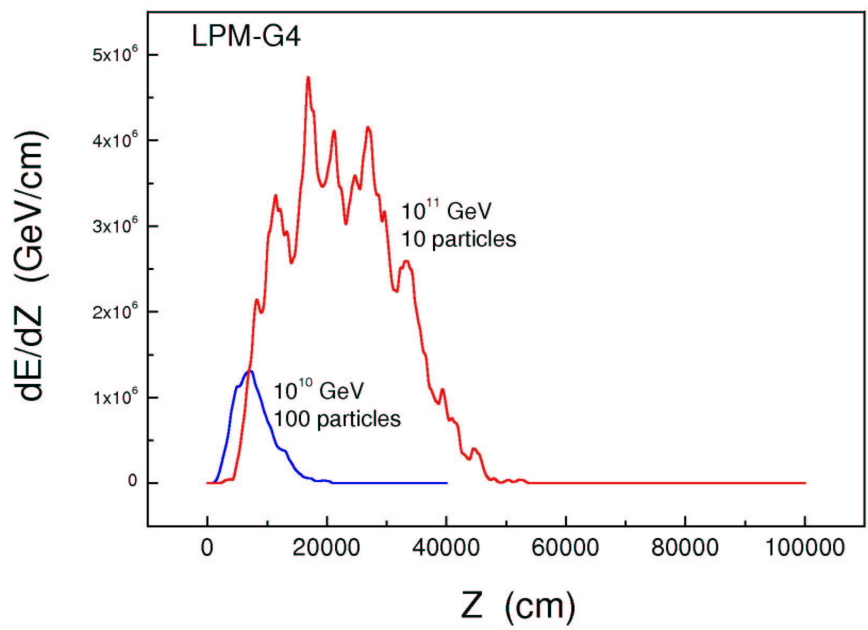

**Fig. 4.**

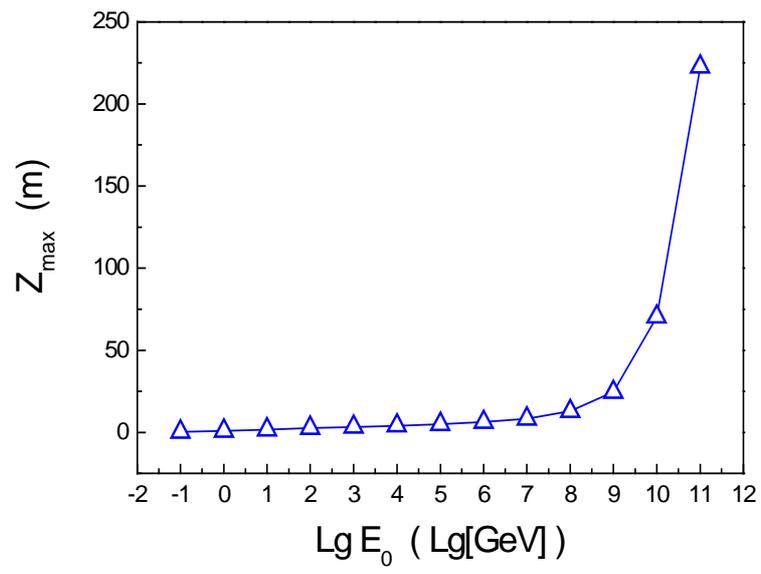

**Fig. 5.**

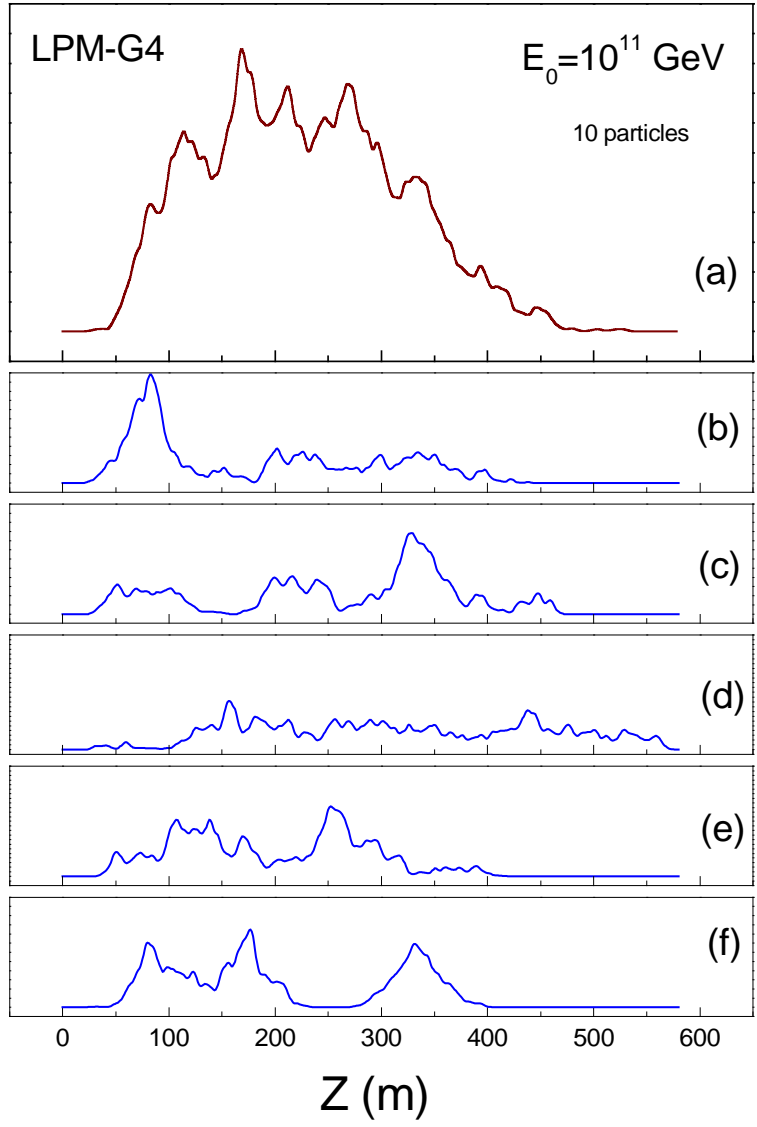

**Fig. 6.**